\documentclass[prd,showpacs,amsmath,amssymb,superscriptaddress,preprintnumbers,nofootinbib,showkeys]{revtex4}

\begin{document}

\newcommand{\nablab}{{\mathop {\rule{0pt}{0pt}{\nabla}}\limits^{\bot}}\rule{0pt}{0pt}}

\title{Is the axionic dark matter an equilibrium system?}

\author{Alexander B. Balakin}
\email{Alexander.Balakin@kpfu.ru} \affiliation{Department of
General Relativity and Gravitation, Institute of Physics, Kazan
Federal University, Kremlevskaya str. 16a, Kazan 420008, Russia}

\author{Amir F. Shakirzyanov}
\email{AmFShakirzyanov@stud.kpfu.ru} \affiliation{Department of
General Relativity and Gravitation, Institute of Physics, Kazan
Federal University, Kremlevskaya str. 16a, Kazan 420008, Russia}

\date{\today}

\begin{abstract}
We consider an axionic dark matter model with a modified periodic potential for the pseudoscalar field in the framework of the axionic extension of the Einstein-aether theory.
The~modified potential is assumed to be equipped by the guiding function, which depends on the expansion scalar constructed as the trace of the covariant derivative of the aether velocity four-vector. The~equilibrium state of the axion field is defined as the state, for which the modified potential itself and its first derivative with respect to the pseudoscalar field are equal to zero.
We apply the developed formalism to the homogeneous isotropic cosmological model, and~find the basic function, which describes the equilibrium state of the axionic dark matter in the expanding Universe.
\end{abstract}
\maketitle

\section{Introduction}

The structure of the cosmic dark matter is one of the most disputable topics in Modern Cosmology~\mbox{\cite{DM1,DM2,DM3,DM4,DM5}}.
There are two opposite points of view in the discussions concerning the dark matter structure. The~first opinion is that this cosmic substratum is formed by a cold massive collisionless and pressureless gas, which participates in the gravitational interaction only. The~diametrically opposite view is that the dark matter can be a system with internal self-interaction and can form the axionic Bose-Einstein condensate \cite{BEC}.

In this article, we consider the models in which the dark matter has the axionic origin, and~the pseudoscalar (axion) field is characterized, first, by the internal self-interaction, second, by the external regulation. To describe the self-interaction, we use the potentials of the pseudoscalar field of two types: the extended periodic and the extended Higgs potentials.
To present an example of the external regulations of the axionic dark matter behavior we work with the axionic extension \cite{ABEAA} of the Einstein-aether theory, which is based on the introduction of the unit time-like vector field $U^i$ interpreted as the aether velocity four-vector \cite{a1,a2,a3,a4,a5}. To link these two trends, we postulate that the potential $V$ of the axion field $\phi$ depends on the scalars, constructed using the covariant derivative of the unit vector field $\nabla_kU^i$. The~formal structure of this potential is the following: $V=V(|\phi|, \Phi_*)$, where~the function describing the basic state of the axion field $\Phi_*$ is not a constant now. It can be presented by a specific function of the guiding scalars, associated with the aether velocity $\Phi_{*}(\Theta, a^2, \sigma^2, \omega^2)$, where~$\Theta=\nabla_kU^k$ is the scalar of expansion of the aether flow; $a^2$ is the square of the acceleration four-vector $a^i$; $\sigma^2$ is the square of the shear tensor $\sigma_{ik}$, and~$\omega^2$ is the square of the vorticity tensor of the aether flow $\omega_{ik}$. In~other words, the~basic state of the axion field is influenced by the spacetime geometry via these four guiding functions.

Why do we think that this model could be interesting for readers? First of all, we recall that the Einstein-aether theory belongs to the class of Modified Theories of Gravity; the theories of this type attract attention, since they successfully explain the observed late-time accelerated  expansion and describe inflation in the early Universe (see, e.g., \cite{MG1,MG2} for details). On the other hand, our work is an attempt to describe the interaction inside the cosmic dark fluid: according to our model, the~axions form the dark matter constituent, and~the dynamic aether plays the role of the dark energy. Clearly, the~interaction of these constituents of the dark fluid is mutual: the aether regulates the state of the axionic dark matter via the modified potential of the pseudoscalar field; in its turn the vector field evolution is influenced by the axionic field. As the result of such internal coupling, the~history of the Universe evolution is expected to be more sophisticated than in the case, when the dark substrata interact only by the gravitational field.

In the works \cite{BG1,BG2,BG3}  we have introduced the concept of the equilibrium state of the pseudoscalar (axion) field. We indicate the value of the pseudoscalar (axion) field $\phi =\phi_{(\rm eq)}$ as the equilibrium one, if~for this  value of the pseudoscalar field the potential $V$ itself and its first derivative $\frac{\partial V}{\partial \phi}$ take zero values. This idea was realized in  \cite{BG1,BG2,BG3} on the example of the axion field distribution in the static spherically symmetric dyon  field.
The goal of this article is to solve the problem of representation of the function $\Phi_{*}$ for the axion field with modified periodic potential in the framework of the cosmological model based on the Einstein-aether-axion theory.

The paper is organized as follows. In~Section II, we recall the basic elements of the mathematical formalism of the Einstein-aether-axion theory. In~Section III we obtain and analyze exact solutions for the equilibrium functions for the spatially isotropic homogeneous Universe using two axion field potentials: the modified Higgs potential and the modified periodic one. Section IV contains conclusions.

\section{Formalism of Axionic Extension of the Einstein-Aether Theory}

\subsection{The Action Functional}

We consider the action functional of the axionic extension of the Einstein-aether theory in the following~form:
\begin{equation}
S_{({\rm EAA)}} =  \int d^4 x \sqrt{{-}g} \ \left\{ \frac{1}{2\kappa}\left[R{+}2\Lambda {+} \lambda (g_{mn}U^m
U^n {-}1 ){+} K^{ab}_{mn} \nabla_a U^m \nabla_b U^n \right] +
\frac{1}{2}\Psi^2_0 \left[V {-} g^{mn}\nabla_m \phi \nabla_n \phi \right] \right\} \,.
\label{total}
\end{equation}

This action functional contains the standard geometrical objects: $g$ is the determinant of the metric $g_{ik}$, $R$ is the Ricci scalar, $\nabla_k$ is the covariant derivative.  As well, $\Lambda$ is the cosmological constant; $U^i$ is the vector field associated with the aether velocity; the dimensionless quantity $\phi$ denotes the pseudoscalar (axion) field. The~parameter $\Psi_0$ is reciprocal
to the constant of the axion-photon coupling $g_{({\rm A} \gamma \gamma)} {=}\frac{1}{\Psi_0}$; the constraint for the constant $g_{({\rm A} \gamma \gamma)}$ is  $g_{({\rm A} \gamma \gamma)}< 1.47 \times 10^{-10} {\rm GeV}^{-1}$. $\lambda$ is the Lagrange multiplier in front of the term, which guarantees that the vector field is normalized by unity. The~constitutive tensor  $K^{ab}_{mn}$ given by the formula
\begin{equation}
K^{ab}_{mn}{=} C_1 g^{ab} g_{mn} {+} C_2 \delta^{a}_{m} \delta^{b}_{n}
{+} C_3 \delta^{a}_{n}\delta^{b}_{m} {+} C_4 U^{a} U^{b}g_{mn} \,,
\label{K}
\end{equation}
contains four coupling constants $C_1$, $C_2$, $C_3$, $C_4$ introduced phenomenologically (see \cite{a1}).
The crucial element of our work is the structure of the potential of the pseudoscalar (axion) field $V$. Standardly, it~is the even function of the axion field $\phi$ only. In~our work the potential $V$ depends on the invariants based on the decomposition of the covariant derivative of the vector field $ \nabla_{i} U_k$:
\begin{equation}
\nabla_i U_k = U_i DU_k + \sigma_{ik} + \omega_{ik} +
\frac{1}{3} \Delta_{ik} \Theta \,. \label{act3}
\end{equation}

The acceleration four-vector $DU^i$, the~symmetric shear tensor $\sigma_{mn}$,  the skew-symmetric vorticity tensor $\omega_{mn}$, and~the expansion scalar $\Theta$ are the irreducible elements
of this decomposition defined as~follows:
$$
DU_k \equiv  U^m \nabla_m U_k \,, \quad \sigma_{ik}
\equiv \frac{1}{2}\left(\nablab_i U_k {+}
\nablab_k U_i \right) {-} \frac{1}{3}\Delta_{ik} \Theta  \,, \quad \Delta^i_k = \delta^i_k - U^iU_k \,,
$$
\begin{equation}
\omega_{ik} \equiv \frac{1}{2} \left(\nablab_i U_k {-} \nablab_k U_i \right) \,, \quad \Theta \equiv \nabla_m U^m \,,
\quad D \equiv U^i \nabla_i \,,  \quad \nablab_i \equiv \Delta_i^k \nabla_k \,.
\label{act4}
\end{equation}

In these terms the  kinetic term takes the form
$$
K^{abmn}(\nabla_a U_m) (\nabla_b U_n) =
$$
\begin{equation}
=(C_1 {+} C_4)DU_k DU^k {+}
(C_1 {+} C_3)\sigma_{ik} \sigma^{ik} + (C_1 {-} C_3)\omega_{ik}
\omega^{ik} {+} \frac13 \left(C_1 {+} 3C_2 {+}C_3 \right) \Theta^2
\,. \label{act5n}
\end{equation}

\subsection{Modifications of the Pseudoscalar Field Potential}

In our work we use two potentials describing the axion field self-interaction.

\subsubsection{Higgs Type Potential}

The first potential has the form of the Higgs potential
\begin{equation}
V(\phi^2, \Phi_{*}) = \frac12 \gamma \left[\phi^2- \Phi^2_* \right]^2 \,,
	\label{Higgs}
	\end{equation}
where $\gamma$ is the phenomenological constant, and~$\Phi_{*}$ is the so-called envelope function, which depends on the scalars $\Theta$, $a^2$, $\sigma^2$, $\omega^2$.
This potential possesses two minima $\phi = \pm \Phi_{*}$ and one maximum at $\phi=0$.

\subsubsection{Periodic Potential of the Axion Field}

Basically, the~axion field representation is characterized by the discrete symmetry $\tilde{\theta} = \frac{\phi}{\ \Phi_*} \to \frac{\phi}{\ \Phi_*}{+} 2\pi$, and~the potential of the axion field inherits this symmetry
\begin{equation}
	V(\phi,\Phi_*) = \frac{m^2_A \Phi^2_{*}}{2\pi^2} \left[1- \cos{\left(\frac{2 \pi \phi}{\Phi_*}\right)} \right] \,.
	\label{potential}
	\end{equation}
	
At $\phi=n \Phi_*$, where $n=0, \pm 1, \pm 2, ...$ this periodic potential has the minima. When $n \to m + \frac12$ we find the maxima of the potential. Near the minimum, when
$\phi \to n \Phi_* + \psi$ and $|\psi|$ is small, $V \to m^2_A \psi^2$, where $m_A$ is the axion rest mass. Mention that the case $\phi =0$ relates to the minimum now, while for the Higgs type potential the case $\phi =0$ describes the maximum.

\subsubsection{Equilibrium State of the Axion Field}

We indicate the state of the axion field as the equilibrium one, $\phi_{(eq)}$, if
 \begin{equation}
	V_{|\phi=\phi_{(eq)}} = 0 \,, \quad \left(\frac{\partial V}{\partial \phi}\right)_{|\phi=\phi_{(eq)}} = 0 \,.
	\label{potential2}
	\end{equation}
	
Clearly, the~Higgs potential defines the stable equilibrium state, when $\phi = \pm \Phi_{*}$. For the periodic potential, we deal with infinite number of the stable equilibrium states at $\phi = n \Phi_*$.

\subsection{Master Equations for the Pseudoscalar, Vector, and~Gravitational Fields}

\subsubsection{Master Equation for the Axion Field}

Variation of the action functional $S_{({\rm EAA)}}$ Equation~(\ref{total}) with respect to pseudoscalar field $\phi$ gives the master equation in the standard form
\begin{equation}
g^{mn} \nabla_m \nabla_n \phi +  \frac12 \frac{\partial V}{\partial \phi} =0 \,.
\label{ax10}
\end{equation}

In the equilibrium state the key equation for the pseudoscalar (axion) field simplifies and takes the~form
\begin{equation}
\nabla_m \nabla^m \Phi_* =0 \,,
\label{ax11}
\end{equation}
i.e., it converts into the equation for the basic state function $\Phi_{*}$, which looks like the Klein–Gordon equation for massless particles. However, we stress that this form of equation corresponds to the nonlinear potential, the~derivative of which vanishes due to the specific self-interaction in the axionic~system.

\subsubsection{Equations for the Unit Dynamic Vector Field}

Variation of the action Equation~(\ref{total}) with
respect to the Lagrange multiplier $\lambda$ yields the equation
\begin{equation}
g_{mn}U^m U^n = 1 \,,
\label{21}
\end{equation}
which is the normalization condition of the time-like
vector field $U^k$. Variation of the functional Equation~(\ref{total}) with respect to
$U^i$ yields the master equation for the aether velocity:
\begin{equation}
\nabla_a {\cal J}^{aj}
 = \lambda \ U^j  + I^j_{(0)} + I^j_{(\rm V)}  \,.
\label{0A1}
\end{equation}

Here ${\cal J}^{aj}$ and
$I^j_{(0)}$ are of the standard form
\begin{equation}
{\cal J}^{a}_{\ j} = K^{ab}_{jn} (\nabla_b U^n)  \,, \quad I^j_{(0)} =  C_4 (DU_m)(\nabla^j U^m) \,.
\label{J7}
\end{equation}

The term $I^j_{(\rm V)}$ depends essentially on the structure of the function $\Phi_*$. In~this work we restrict the model by the condition that
this basic state function depends on the expansion scalar only, $\Phi_* = \Phi_*(\Theta)$. This restriction is motivated by the idea to consider the application to the cosmological model of the Friedmann type only, for which $a^2=0$, $\sigma^2=0$, $\omega^2=0$. This ansatz yields
\begin{equation}
I^j_{(\rm V)} =  - \frac12 \kappa \Psi^2_0 \nabla^j \left[\frac{\partial V}{\partial \Phi_{*}} \frac{d \Phi_*}{d \Theta} \right] \,.
\label{J2}
\end{equation}

The Lagrange multiplier $\lambda$ can be obtained standardly as
\begin{equation}
\lambda =  U_j \left[\nabla_a {\cal J}^{aj}- I^j_{(0)} - I^j_{(\rm V)}\right]  \,.  \label{0A309}
\end{equation}

\subsubsection{Equations for the Gravitational Field}

The variation of the action Equation~(\ref{total}) with respect to the metric
$g^{ik}$ yields the equations
\begin{equation}
R_{ik} - \frac{1}{2} R \ g_{ik}
=  \Lambda g_{ik}  + T^{(\rm U)}_{ik} +
\kappa T^{({\rm A})}_{ik} + \kappa T^{(\rm V)}_{ik} \,. \label{0Ein1}
\end{equation}

Here
$$
T^{(\rm U)}_{ik} =
\frac12 g_{ik} \ K^{abmn} \nabla_a U_m \nabla_b U_n{+} \lambda U_iU_k {+}
\nabla^m \left[U_{(i}{\cal J}_{k)m} {-}
{\cal J}_{m(i}U_{k)} {-}
{\cal J}_{(ik)} U_m\right]+
$$
\begin{equation}
+C_1\left[(\nabla_mU_i)(\nabla^m U_k) {-}
(\nabla_i U_m )(\nabla_k U^m) \right] +
C_4 DU_i DU_k  \,.
\label{5Ein1}
\end{equation}

The symbol $p_{(i} q_{k)}$
denotes symmetrization.
The quantity
\begin{equation}
T^{({\rm A})}_{ik} = \Psi^2_0 \left[\nabla_i \phi \nabla_k \phi
+\frac12 g_{ik}\left(V {-} \nabla_n \phi \nabla^n \phi \right) \right]
\label{qq1}
\end{equation}
is the extended stress-energy tensor of the pseudoscalar field.
The last term
\begin{equation}
 T^{(\rm V)}_{ik} = - \frac12 \Psi^2_0 g_{ik} \nabla_j \left[U^j \frac{\partial V}{\partial \Phi_{*}} \frac{d \Phi_*}{d \Theta}  \right]
\label{int}
\end{equation}
appeared as the result of variation of the expansion scalar $\Theta$, the~argument of the function $\Phi_*$ included into the potential $V(|\phi|, \Phi_{*}(\Theta))$.

\section{Application to the Friedmann-Type Cosmological Model}

\subsection{Reduced Master Equations}

 Let us consider the master equations for the pseudoscalar, vector and gravitational field for the cosmological model with the metric
 \begin{equation}
ds^2 = dt^2 - a^2(t)[dx^2+dy^2+dz^2] \,.
\label{App1}
\end{equation}

As usual, $a(t)$ is the scale factor, $H(t){=}\frac{\dot{a}}{a}$ is the Hubble function, and~the dot denotes the derivative with respect to cosmological time $t$.
Our ansatz is that the pseudoscalar and unit dynamic vector fields inherit the chosen symmetry, so that the state  functions  depend on the cosmological time only, $\phi(t)$~and $U^i(t)$, and~the velocity four-vector is of the form $U^i = \delta^i_0$, which guarantees the equivalence of all spatial directions in the Friedmann world. For this symmetry, the~covariant derivative $\nabla_i U_k$  is characterized by vanishing acceleration four-vector, shear and vorticity tensors:
\begin{equation}
DU^i = 0 \,, \quad \sigma_{mn}=0\,, \quad \omega_{mn}=0 \,.
\label{App2}
\end{equation}

The expansion scalar is now proportional to the Hubble function $\Theta = 3H(t)$, and~the covariant derivative can be written as follows:
\begin{equation}
 \nabla_i U_k =  \Delta_{ik} \ H(t) \,.
\label{App3}
\end{equation}

Let us consider now the evolutionary equations for the unit vector field, for the pseudoscalar and gravitational fields.

\subsubsection{Reduced Equations for the Unit Vector Field}

Keeping in mind the symmetry of the model under consideration, we see that
the four-vector $I^j_{(0)}$ vanishes, the~tensor ${\cal J}^{aj}$ reduces to
\begin{equation}
{\cal J}^{aj} {=} H \left[\Delta^{aj}\left(C_1{+}3C_2{+}C_3 \right) {+} 3C_2 U^a U^j \right] \,,
\label{9J2}
\end{equation}
and the term $I^j_{(\rm V)}$ takes the form
\begin{equation}
I^j_{(\rm V)} = \frac16 \kappa \Psi^2_0 U^j \frac{d}{dt}\left[\frac{\partial V}{\partial \Phi_{*}} \frac{d \Phi_*}{d H} \right] \,.
\label{J29}
\end{equation}

Clearly, the~divergence $\nabla_a {\cal J}^{aj}$ is proportional to the velocity four-vector, thus,
three of four evolutionary equations for the unit vector field are satisfied identically, and~the last equation  defines the Lagrange multiplier:
\begin{equation}
\lambda(t) =
-3H^2 \left(C_1+C_3 \right) + 3C_2 \dot{H} +
\frac16 \kappa \Psi^2_0 \frac{d}{dt} \left[\frac{\partial V}{\partial \Phi_{*}} \frac{d \Phi_*}{d H} \right] \,.
\label{App11}
\end{equation}

In other words, the~equations for the unit vector field are solved, and~the obtained function $\lambda(t)$ will be now inserted into the equations for the gravity field.

\subsubsection{Reduced Equation for the Pseudoscalar (Axion) Field}

The reduced evolutionary equation for the axion field is now of the form
\begin{equation}
\ddot{\phi} + 3H \dot{\phi} +  \frac12 \frac{\partial V}{\partial \phi} =0 \,.
\label{App125}
\end{equation}

For the equilibrium state this equation reduces to $\ddot{\Phi}_* + 3H \dot{\Phi}_* =0$ for both cases: for the Higgs type potential (when $\phi=\pm \Phi_*$), and~for the periodic potential (when $\phi= n \Phi_*$).

\subsubsection{The Key Equation for the Gravitational Field}

For the Friedmann-type model only one equation for the gravity field is independent; we indicate it as the key equation. It can be written in the following form:
\begin{equation}
 3H^2 \left[1+ \frac12 \left(C_1{+}3C_2{+}C_3 \right) \right] {-} \Lambda
= \frac{\kappa \Psi^2_0}{2} \left[ {\dot{\phi}}^2 + V - H \left(\frac{\partial V}{\partial \Phi_{*}} \right) \left(\frac{d \Phi_*}{d H}\right) \right]
\,.
\label{App186}
\end{equation}

Other gravity field equations and the conservation law are the consequences of the key equation, equations for the unit vector field and equation for the axion field.

\subsection{Equilibrium State with Vanishing Axion Field}

The first test model is based on the ansatz that in the basic equilibrium state the axion field is vanishing, $\phi=0$.
What is the envelope function $\Phi_*$ in this case. The~predictions obtained for the modified periodic and for the Higgs type potentials are different.

\subsubsection{The Solution with the Periodic Potential}

When $\phi=0$ we see that for the modified periodic potential of the axion field
\begin{equation}
V(0, \Phi_{*}) = 0  \,, \quad \left(\frac{\partial V}{\partial \phi}\right)_{|\phi =0} =0 \,, \quad \left(\frac{\partial V}{\partial \Phi_*}\right)_{|\phi =0} = 0 \,.
	\label{Higgs99}
	\end{equation}
	
The equation for the axion field is satisfied identically, and~the key equation for the gravity field takes the form
\begin{equation}
3H^2 \left[1+ \frac12 \left(C_1{+}3C_2{+}C_3 \right) \right] {-} \Lambda  = 0 \,.
\label{H9}
	\end{equation}
	
The appropriate solution to this equation is the solution of the de Sitter type
\begin{equation}
H(t) = H_{\infty} \,, \quad H_{\infty} \equiv \sqrt{\frac{\Lambda}{3\Gamma}} \,, \quad \Gamma \equiv 1 + \frac12 \left(C_1{+}3C_2{+}C_3 \right) \,.
\label{H99}
	\end{equation}
	
In this case, the~envelope function $\Phi_{*}(H)$ cannot be found, but can be postulated, e.g., as a~constant.

\subsubsection{The Solution with the Higgs Type Potential}

When $\phi=0$, for the Higgs type potential we obtain the following auxiliary formulas:
\begin{equation}
V(0, \Phi_{*}) = \frac12 \gamma \Phi^4_*  \,, \quad \left(\frac{\partial V}{\partial \phi}\right)_{|\phi =0} =0 \,, \quad \left(\frac{\partial V}{\partial \Phi_*}\right)_{|\phi =0} = 2 \gamma \Phi^3_* \,.
	\label{Higgs999}
	\end{equation}
	
Again, the~equation for the axion field is satisfied identically, but the key equation for the gravity field has more complicated form
\begin{equation}
3H^2 \left[1+ \frac12 \left(C_1{+}3C_2{+}C_3 \right) \right] {-} \Lambda  = - \frac{\kappa \gamma \Psi^2_0}{4} H^2 \frac{d}{dH}\left( \frac{\Phi^4_*}{H}\right) \,.
\label{Hi9}
	\end{equation}
	
The formal solution to this equation is
\begin{equation}
\Phi_{*}(t) = \left\{\Phi^4_{*}(t_0) \frac{H(t)}{H(t_0)} - \frac{12 \Gamma}{\kappa \gamma \Psi^2_0} \left[H(t)-H(t_0) \right]\left[H(t)- \frac{H^2_{\infty}}{H(t_0)}  \right]\right\}^{\frac14} \,,
\label{Hig9}
	\end{equation}
but now the Hubble function is arbitrary. If we prefer again to have a constant value   $\Phi_{*}(t) = const$, we can put $H=const$ and find that
\begin{equation}
H^2   = H^2_{\infty}  +  \frac{\kappa \gamma \Psi^2_0}{12 \Gamma} \Phi^4_* \,.
\label{mH}
	\end{equation}
	
The difference between these two model potentials can be explained as follows. The~value $\phi=0$ relates to the minimum of the periodic potential, so, this state is the stable equilibrium one; as for the Higgs type potential, the~value $\phi=0$ corresponds to the maximum, and~thus to the instable state.

\subsection{Equilibrium State with Non-Vanishing Axion Field}

We consider now the key equation for the gravity field
$$
\frac{2}{\kappa \Psi^2_0 } \left\{3H^2 \left[1+ \frac12 \left(C_1{+}3C_2{+}C_3 \right) \right] {-} \Lambda \right\} =
$$
\begin{equation}
=  {\dot{\phi}}^2
+
\frac{m^2_A }{2 \pi^2}   \left\{-H^2
\left[1- \cos{\left(\frac{2 \pi \phi}{\Phi_*}\right)} \right]\frac{d}{dH}\left(\frac{\Phi^2_{*}}{H} \right) + 2 \pi \phi \sin{\left(\frac{2 \pi \phi}{\Phi_*}\right)} H \frac{d \Phi_*}{d H}
 \right\}
\,,
\label{2App186}
\end{equation}
and put $\phi = n \Phi_*$ with  $n \neq 0$. This equation takes the form
\begin{equation}
3H^2 \left[1+ \frac12 \left(C_1{+}3C_2{+}C_3 \right) \right] {-} \Lambda  = \frac{\kappa \Psi^2_0 n^2}{2} {\dot{\Phi}_*}^2 \,.
\label{A186}
\end{equation}

The equation for the axion field is now nontrivial
\begin{equation}
\ddot{\Phi}_{*} + 3H \dot{\Phi}_{*} = 0 \,.
\label{HH1}
\end{equation}

The first integral of this equation is
\begin{equation}
\dot{\Phi}_{*}(t) = \dot{\Phi}_{*}(t_0) \left(\frac{a(t)}{a(t_0)} \right)^{-3} \,,
\label{2HH1}
\end{equation}
and its structure advises to use the auxiliary variable
$x \equiv \frac{a(t)}{a(t_0)} $ with the differentiation according to the rule
$ \dot \Phi_{*} = xH(x)\frac{d\Phi_*}{dx}$.
In these terms we obtain the following system of equations:
\begin{equation}
H^2(x)  = H^2_{\infty} +  \frac{\kappa \Psi^2_0}{6\Gamma}  n^2  \dot{\Phi}^2_{*}(t_0) x^{-6}
\,,
\label{HjH2}
\end{equation}
\begin{equation}
H \frac{d}{dx}\Phi_{*}(x) = \dot{\Phi}_{*}(t_0) x^{-4}\,.
\label{3HH1}
\end{equation}

To solve this system, we find, first, the~quantity $x$ as the function of the Hubble function
\begin{equation}
x= \left[\frac{\kappa \Psi^2_0 n^2  \dot{\Phi}^2_{*}(t_0)}{6\Gamma (H^2{-}H^2_{\infty})} \right]^{\frac16} \,,
\label{HH17}
\end{equation}
second, using this formula we replace $dx$ with $dH$, and~third, we obtain the equation
\begin{equation}
\frac{d\Phi_*}{dH} = - \sqrt{\frac{2\Gamma}{3\kappa n^2 \Psi^2_0(H^2{-}H^2_{\infty})}} \,,
\label{HH15}
\end{equation}
the solution to which is of the form
\begin{equation}
	\Phi_*(H)= \Phi_{\infty} \mp  \sqrt{\frac{2\Gamma}{3 n^2 \kappa \Psi^2_0}} \ ln{\left[ \frac{\sqrt{H^2{-}H^2_{\infty}}{+}H}{H_{\infty}}\right]}\,.
\label{HH20}
\end{equation}

In the limiting case $n \to 0$ we can obtain the finite value for the envelope function if and only if the argument of the logarithm is equal to one; it is possible when $H=H_{\infty}$. In~other words, in~this case the spacetime is of the de Sitter type, and~the envelope function is equal to unknown constant.

To find the scale factor $a(t)$ we use the standard relationship
\begin{equation}
t-t_0 = \int^{\frac{a(t)}{a(t_0)}}_1 \frac{dx}{x H(x)} \,, \quad H(x)= \sqrt{H^2_{\infty} + x^{-6} \frac{\kappa \Psi^2_0 n^2 \dot{\Phi}^2_*(t_0)}{6\Gamma}} \,.
\label{HH77}
\end{equation}

Direct integration gives the following formulas for the scale factor and Hubble function:
\begin{equation}
a(t)= a(t_0) \left\{\cosh{\left[3H_{\infty}(t-t_0)\right]}  + \sqrt{1+\frac{\kappa \Psi^2_0 n^2 \dot{\Phi}^2_*(t_0)}{2 \Lambda}} \sinh{\left[3H_{\infty}(t-t_0)\right]}\right\}^{\frac13} \,,
\label{H66}
\end{equation}
\begin{equation}
H(t) = H_{\infty} \  \left\{\frac{\sinh{\left[3H_{\infty}(t-t_0)\right]}  + \sqrt{1+\frac{\kappa \Psi^2_0 n^2 \dot{\Phi}^2_*(t_0)}{2 \Lambda}} \cosh{\left[3H_{\infty}(t-t_0)\right]}}{\cosh{\left[3H_{\infty}(t-t_0)\right]}  + \sqrt{1+\frac{\kappa \Psi^2_0 n^2 \dot{\Phi}^2_*(t_0)}{2 \Lambda}} \sinh{\left[3H_{\infty}(t-t_0)\right]}}\right\} \,.
\label{H33}
\end{equation}

In the asymptotic regime the obtained solutions have quasi-de Sitter asymptotes
\begin{equation}
 a(t \to \infty) \propto e^{H_{\infty} t} \,, \quad H(t \to \infty) \to H_{\infty}  \,.
\label{HH71}
\end{equation}

It is interesting to mention that in the work \cite{BSh} we obtained the similar formulas for the model with the Higgs type potential; e.g., the~formula for $\Phi_*$ can be extracted from Equation~(\ref{HH20}), if to put $n=1$. We~think that it is a good signal that Equation~(\ref{HH20}) gives some universal basic state function for description of the equilibrium axionic dark matter in the isotropic spatially homogeneous spacetime.

\section{Conclusions}

In the work \cite{BSh} we studied in detail the cosmological model, in~which the dynamic aether plays the role of dark energy, and~the dark matter is described by the axionic system with the modified Higgs  potential describing self-interaction of the $\phi^4$ type. In~the presented work we extend the model by introduction of the modified periodic potential of the pseudoscalar (axion) field. This extended model predicts that there is an infinite number of equilibrium states for the axion field, which correspond to the minima of the periodic potential. We have found the so-called envelope functions, which describes the set of basic equilibrium states of the axionic dark matter (see Equation~(\ref{HH20})). This equilibrium functions depend on the Hubble function, and~this fact allows us to interpret the interaction between the axionic dark matter and the dynamic aether as the process of regulation realized by the aether.
As in the model studied in \cite{BSh}, the~isotropic spatially homogeneous spacetime behaves asymptotically as the de Sitter type Universe, for which the Hubble function inherits the information about the cosmological constant $\Lambda$ and  coupling constants $C_1$, $C_2$,$C_3$ describing the dynamic aether (see Equation~(\ref{H99})).

The first conclusion is that one function from the set of basic equilibrium functions Equation~(\ref{HH20}) can be considered to be a {\it universal guiding state function}; however, the~question of what is the number $n$, which corresponds to this universal function, remains open.

The second conclusion is that there exists the backreaction of the axionic dark matter on the dark energy associated with the aether; this fact can be illustrated by the terms $I^j_{(\rm V)}$ (see Equation~(\ref{J29})) and  by the Lagrange multiplier $\lambda(t)$ (see Equation~(\ref{App11})), which enter the master equations for the unit vector field.

The third conclusion is that the interactions inside the dark fluid modify the rate of the Universe evolution. This sentence can be illustrated by the formulas for the scale factor Equation~(\ref{H66}) and for the Hubble function Equation~(\ref{H33}); in both formulas there are two guiding parameters, which describe the rate of evolution: the  first constant $\Psi^2_0 \dot{\Phi}^2_*(t_0)$ is associated with the axionic dark matter, and~the second constant $H_{\infty}$ (see Equation~(\ref{H99})) is associated with the dark energy. These results are analogous in many aspects to the results obtained in the works \cite{MG3,MG4}.

\acknowledgments{The work was supported by Russian Foundation for Basic Research (Project No. 20-02-00280), and, partially, by the Program of Competitive Growth of Kazan Federal University.}

\end{document}